\documentclass[5p,times,authoryear]{elsarticle}

\usepackage{ecrc}

\volume{00}

\firstpage{1}

\journalname{Astronomy and Computing}

\runauth{O. Creaner, K. Nolan, J. Walsh \& E. Hickey}

\jid{astron.comput.}

\jnltitlelogo{Astronomy and Computing}

\usepackage{amssymb}

\usepackage[figuresright]{rotating}

\usepackage[table]{xcolor}

\usepackage{makecell}

\begin{document}

\begin{frontmatter}



\dochead{}

\title{The Locus Algorithm III: A Grid Computing system to generate catalogues of optimised pointings for Differential Photometry}


\author[ITTD,DIAS,LBNL]{Ois\'{i}n Creaner\corref{cor1}}
\cortext[cor1]{Corresponding author}
\ead{creanero@gmail.com, oocreaner@lbl.gov}

\author[ITTD]{Kevin Nolan}
\ead{kevin.nolan@tudublin.ie}

\author[GI]{John Walsh}

\author[ITTD]{Eugene Hickey}
\ead{eugene.hickey@tudublin.ie}

\address[ITTD]{Technological University Dublin, Tallaght Campus, Dublin 24, Ireland}
\address[DIAS]{Dublin Institute for Advanced Studies, 31 Fitzwilliam Place, Dublin 2, Ireland}
\address[GI]{Grid-Ireland, School of Computer Science and Statistics, Trinity College, Dublin 2, Ireland}
\address[LBNL]{Lawrence Berkeley National Laboratory, 1 Cyclotron Road, Berkeley, California, USA}

\begin{abstract}
This paper discusses the hardware and software components of the Grid Computing system used to implement the Locus Algorithm to identify optimum pointings for differential photometry of 61,662,376 stars and 23,799 quasars. The scale of the data, together with initial operational assessments demanded a High Performance Computing (HPC) system to complete the data analysis.  Grid computing was chosen as the HPC solution as the optimum choice available within this project.  The physical and logical structure of the National Grid computing Infrastructure informed the approach that was taken.  That approach was one of layered separation of the different project components to enable maximum flexibility and extensibility.

\end{abstract}


\end{frontmatter}


\section{Introduction}
\label{Introduction}
The Locus Algorithm, first described by \citet{creaner2010large} and explained in full in \citet{locuspaper} is an algorithm that changes the position of the field of view (FoV) to provide optimised conditions for differential photometry, given a target and telescope parameters. It works by translating the FoV on a North-South/East West basis such that the maximum number and quality of reference stars are included in the FoV while ensuring that the target remains in the FoV.  A software system which harnesses this algorithm was developed as shown in \citet{creaner2016thesis} and detailed in \citet{locus_software_paper}.  This system was used to develop two catalogues of pointings based on the Sloan Digital Sky Survey (SDSS) catalogue \citep{schneider2007sloan,abazajian2009seventh}. The first catalogue started with 40,000 quasars from the  as input targets to generate pointings for 23,779 quasars as discussed in \citet{quasarpaper} and with 357,175,411 point sources from SDSS as inputs to produce pointings for 61,662,376 stars for use as candidates for exoplanet observation as shown in \citet{ZenodoXOPCatalogue}.  It is this software system which is described in this paper.  The performance metrics of this system are discussed in \cite{grid_system_paper}.

In principle, the analysis of each target can be carried out independently.  Thus, the analysis lends itself well to parallelisation. The strategy for parallelisation of this system is dictated in part by the hardware system used, but within that framework there are several considerations which are discussed below.  

This paper discusses the conceptual design of this system at several levels: The need for and choice of a High Performance Computing (HPC) system; the structure of the Grid system chosen, and the implications that has on the design of the system; the  approach taken in the creation and implementation of Grid jobs; and finally the design of the software and data structures used to execute those jobs.  A brief discussion of the practical issues which arose during the implementation of this system is also given.

This paper forms part of a series of papers.  The first paper in this series, \citet{locus_software_paper} discusses the design of the software system used in this project when used on a standalone system such as the test system and largely ignores the implications on the design of using grid computing.  This paper, the second, focusses on the grid-specific design considerations and implementation.  The third, \citet{grid_metrics_paper}, provides an assessment of the performance metrics of the grid system implemented, and discusses the implications of this for future similar systems.

\section{High Performance Computing Considerations and Options}
\label{Considerations}
During initial testing, the software for the main data pipeline ran for between 0.25-1.0 seconds with an arbitrary sample of individual test targets on a single computer.  Given a potential maximum of 357,175,411 point sources in SDSS DR7 \citep{abazajian2009seventh} for the generation of pointings for exoplanet candidates, a potential runtime of between  2.8 and 11 years was predicted for the generation of that catalogue  \citep{ZenodoXOPCatalogue}.  This showed the necessiy of a High Performance Computing system to enable this analysis to be completed in a practical timescale. Close inspection of these tests also showed that data I/O operations dominated the runtime over algorithmic processing time.

      \begin{figure}[!htb]
        \center{\includegraphics[width=.47\textwidth]
        {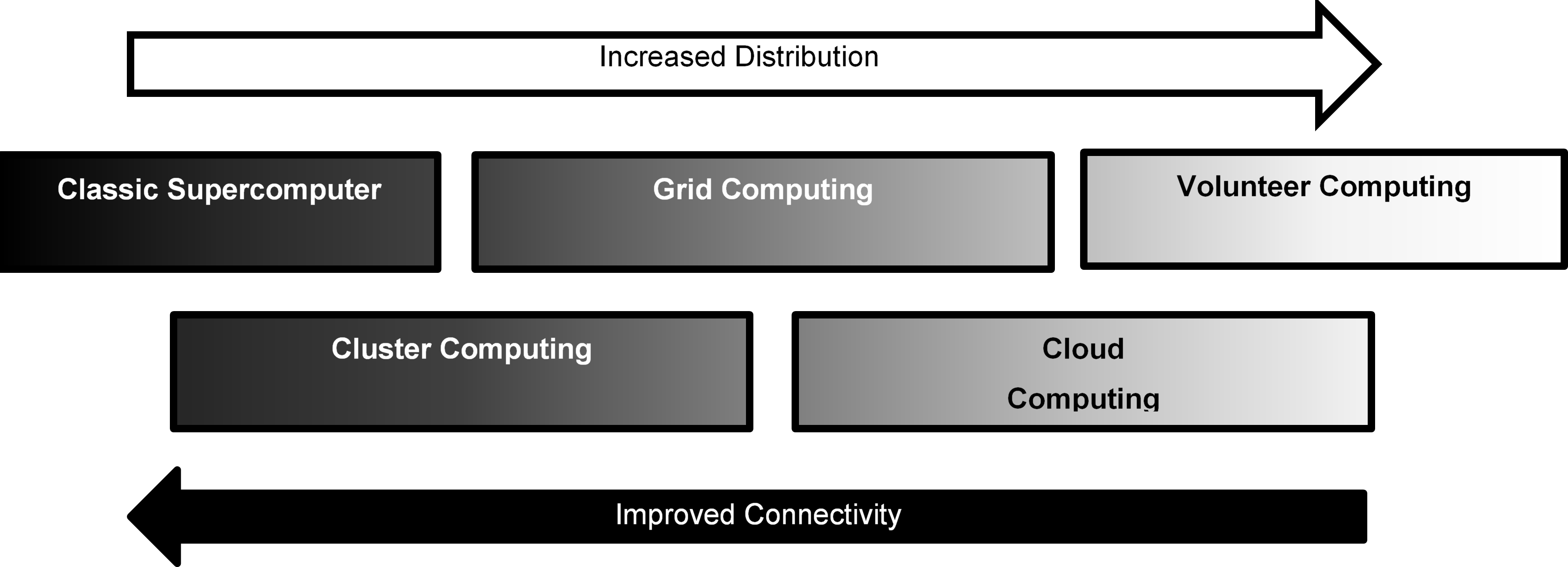}}
        \caption{\label{fig:distribution} Spectrum of High Performance Distributed computing.  Copied from \citet{creaner2016thesis}}
      \end{figure}

A variety of methods exist for increasing the performance of a computing solution beyond that of a standard device - in general, these depend on the strategy of parallel computing to complete the computational challenge \citep{dongarra2006overview}.  By dividing the processing of the data into a number of processing units, and assigning each of these to a separate processor, it becomes possible to complete those analyses simultaneously, and thus reduce the overall runtime \citep{dongarra2006overview}.
      
As shown in Figure \ref{fig:distribution}, these solutions range from Classic Supercomputers: dedicated hardware running specialised, purpose built software with extremely high degree of connectivity between compute and storage elements \citep{dimitrijevic2008virtual}; to volunteer computing systems such as \texttt{SETI@Home} and \texttt{Folding@home} where private users make computing time available on their standard home or office computer for data analysis purposes \citep{beberg2009folding}.  In this project, the former was determined to be unsuitable due to the highly specialised requirements, while the latter would require the recruitment of volunteer contributors beyond the scope of the project \citep{creaner2016thesis}.  This left Cluster, Grid and Cloud computing as viable options for a High Performance Computing solution.

Each of these three paradigms incorporates computing elements which can run software similar or identical to that used for standard devices with a system to manage the distribution of data and processing \citep{dimitrijevic2008virtual,coghlan2005grid}.  Due to the available resources, grid computing was selected as the preferred system for this project.

A grid computing system consists of a collection of similar devices known as \textit{Worker Nodes} (WN) running standard UNIX operating systems.  These WNs can be located at one or more locations and are connected by a Wide Area Network (WAN).  The grid is managed by \textit{middleware}  referred to as the Grid Management System (GMS.) The GMS supports interactions between the WN and shared resources such as long term storage \citep{coghlan2005grid}.  The National Grid computing Infrastructure (NGI) managed by Grid-Ireland was the system used for this project, which used a middleware system known as \texttt{gLite} \citep{coghlan2005grid}. 
\section{NGI and \texttt{gLite} grid operations}
\label{gridoperations}
      \begin{figure}[!htb]
        \center{\includegraphics[width=0.47\textwidth]
        {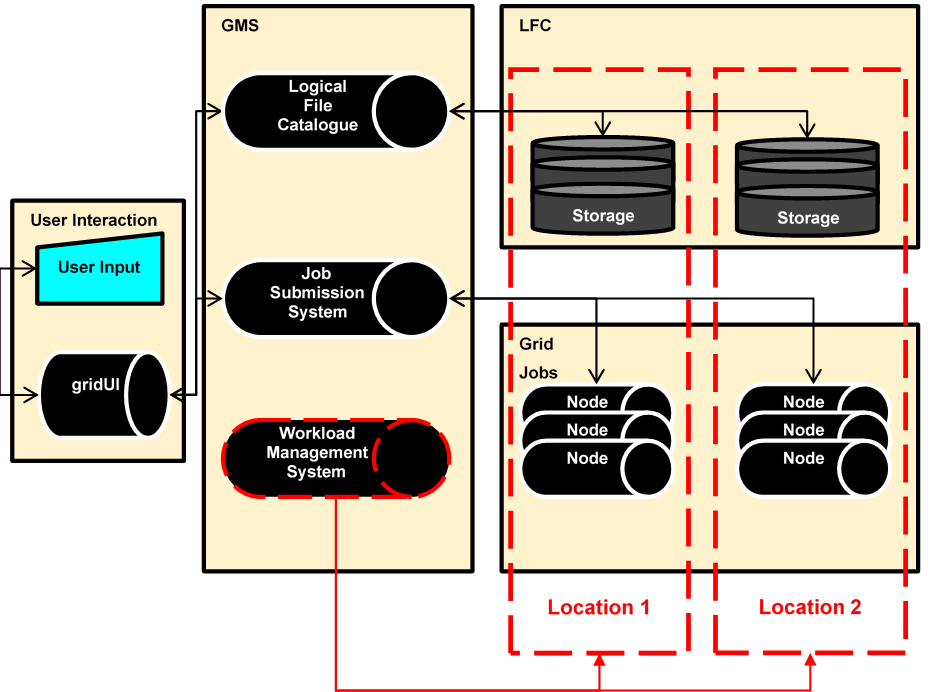}}
        \caption{\label{fig:grid_concept} Conceptual structure of NGI from the user perspective.  Copied from \citet{creaner2016thesis}.}
      \end{figure}

As shown in Figure \ref{fig:grid_concept}, this system conceptually separates the operation of grid jobs into three elements.  The user interface (known as \texttt{gridUI}, see Subsection \ref{gridui}); the data storage, known as the Logical File Catalogue (\texttt{LFC}, see \ref{LFC}) and the grid jobs, which are submitted to the Job Submission System (JSS, see \ref{JSS}) which accepts grid jobs and distributes individual grid jobs among worker nodes \citep{glite}.  The LFC and the JSS together with the Workload Management System (WMS, see \ref{WMS}) form the GMS.

\subsection{gridui}
\label{gridui}
Access to the grid was provided by means of a dedicated computer called \texttt{gridUI} (grid user interface), which acted as a gateway to the grid, located at ITTD.  This computer used the same operating system and shell, and had the same access to the grid as a grid node and as such was ideal for testing and debugging grid software.  In addition, it was on gridUI that grid jobs were generated, and from which jobs were submitted and monitored, and finally to which results were downloaded when needed.

\subsection{LFC}
\label{LFC}
The long-term, large-scale distributed file stoage system used in NGI is the \texttt{LFC} \citep{coghlan2005grid}.  The \texttt{LFC} system consists of three structural components \citep{glite}.  

The first component is a catalogue of Logical File Names (\texttt{LFN}) linked to their corresponding Globally Unique Identifiers (\texttt{GUID}) and Storage Universal Resource Locator (\texttt{SURL}) \citep{gridworkshop}.  \texttt{LFN}s are designed to be human-readable and are structured to mimic a standard UNIX File System (UFS).  \texttt{GUID}s are a unique string of the form guid uniquestring and are used as a primary key for each file in the catalogue. \texttt{SURL}s refer to the location where the file is physically stored.  

The physical storage referred to by the \texttt{SURL} is the second element of the \texttt{LFC} suite \citep{gridworkshop,glite}.  Unless the user specifies otherwise, the physical location of files is assigned automatically by the GMS and is distributed across the various locations of the grid system.  

Data in the \texttt{LFC} cannot be directly accessed, but rather must be accessed by means of the final element of the \texttt{LFC} – the set of gLite middleware commands used to interact with the data in the \texttt{LFC} \citep{glite}.  These commands are designed to mimic the UNIX shell commands.  For example, the gLite command \texttt{lfc-ls} will list the contents of a directory specified by its \texttt{LFN} in the same way that the \texttt{ls} UNIX command will list the contents of a directory in the UNIX file system \citep{glite}.

\subsection{JSS}
\label{JSS}
Similarly, the user does not interact with the grid nodes directly.  Work is instead assigned to the nodes by the JSS \citep{glite}.
Users submit individual tasks to the JSS by means of grid jobs.  Grid jobs are each defined by a single file written in the Job Description Language (JDL).  \texttt{.jdl} files consist of a series of name-value pairs which fully define the task assigned to the grid and specify any requirements that that task may have such as available storage, memory or permitted job time \citep{glite}.  

A \texttt{.jdl} file must include the path to an executable file which will be copied from \texttt{gridUI} to the WN and executed on the WN \citep{glite}.  Additional requirements may include data files that are required for the job \citep{glite}.  Unless the user specifies otherwise in the job, the physical location of the worker node is selected automatically by the Workload Management System (WMS).   The WMS is designed to handle load balancing and job scheduling and thus ensure that WNs are never left idle. 
 
Jobs are submitted to the GMS using \texttt{gLite} commands.  When these commands are executed, the JDL file is sent to the JSS, and the executable file, together with any data files specified in the job, are uploaded to the GMS \citep{glite,gridworkshop}.  These files are then downloaded to the working directory of the WN to which the job is assigned.  This is an essential consideration for the design of the executable software.
  
Further \texttt{gLite} commands exist for the monitoring and management of grid jobs \citep{glite,gridworkshop}. Grid jobs, once submitted to the WMS are given a status to indicate their progress from ``submitted'' to ``cleared'' which may be monitored by the user manually, or automatically as used in this project and discussed in Section \ref{jobmanagment} \citep{glite}.  

Grid jobs may fail to complete for a variety of reasons, including grid problems, such as node crashes or software problems such as memory leaks or missing data.  The GMS does not automatically resubmit failed jobs, instead requiring that the user monitor the progress of jobs, and resubmit them if appropriate.  Checkpointing may be implemented by the user using the LFC and suitable scripting techniques if desired, but is not part of the default system \citep{glite}.

\subsection{WMS}
\label{WMS}
The NGI was a system physically distributed at several widely separated sites connected via WAN.  The Workload Management System monitored the storage and compute elements at these locations, and distributed the workload appropriately.  Unless the use of specific locations was specified by the the user either directly or by specifying distinctive attributes of the nodes, the WMS would automatically select a location for LFC storage or jobs submitted through JSS.  Users did not need to specify locations.

\section{System Design}
\label{Design}

\begin{figure}[!htb]
        \center{\includegraphics[width=0.47\textwidth]
        {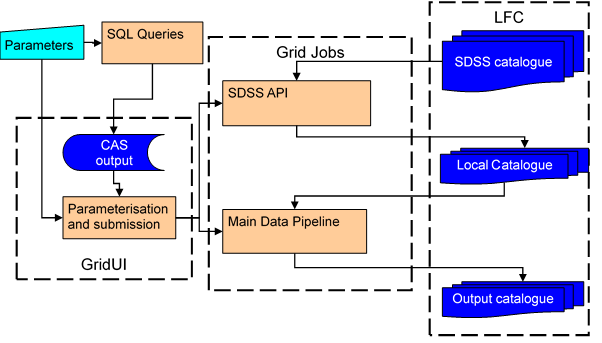}}
        \caption{\label{fig:grid_jobs} Conceptual structure of grid jobs as used in this project.  Copied from \citet{creaner2016thesis}.}
      \end{figure}

The system as implemented used these structural elements as shown in Figure \ref{fig:grid_jobs}.  First the SDSS file catalogue was downloaded to the LFC from the SDSS Data Archive Server (DAS) \citep{SDSSDAS}.  SQL queries to the SDSS Catalogue Archive Server (CAS) \citep{SDSSCAS} were used together with user parameters to parameterise and submit grid jobs through GridUI.  There were two main elements, the SDSS API and the Main Data Pipeline, each of which would be completed by submitting many grid jobs.  

The SDSS API accessed the SDSS Catalogue and separated out a catalogue that met the ``clean sample of point sources'' criteria provided by SDSS \citep{SDSScleansample}.  These sources were stored in the Local Catalogue on the LFC for use as references for all targets and as candidate targets in the generation of the Exoplanet Catalogue \citep{ZenodoXOPCatalogue}.

The main data pipeline implemented the Locus Algorithm, and when provided with target lists of targets from the CAS, was used to produce catalogues of outputs of the optimum pointing for differential photometry of quasars and exoplanet host candidates.

Each of these grid elements was constrained by parameters input by the user, and managed by a suite of bespoke grid management scripts written in Bash, as shown in Figure \ref{fig:grid_jobs}.  These scripts received their parameters from the command line, and based on this input divided the data into appropriately sized units.  These units were designed to be processed on a single node within the time limits set by the grid provider: each such unit constituted one grid job.  The management system, illustrated in Figure \ref{fig:grid_management} would then submit those jobs in a controlled manner to allow the job submission system to handle the data throughput, up to a maximum number of simultaneous jobs again defined by the grid provider.

\section{Job Management}
\label{jobmanagment}

\begin{figure}[!htb]
        \center{\includegraphics[width=0.5\textwidth]
        {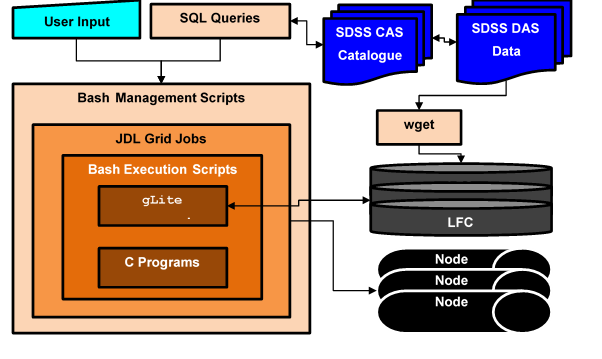}}
        \caption{\label{fig:grid_management} The management of a grid job.  Copied from \citet{creaner2016thesis}}
      \end{figure}

As shown in Figure \ref{fig:grid_management}, each grid job, defined in JDL (the Job Description Language) consisted of a Bash script to execute and a set of parameters to apply to that script which allowed each job to access a different unit of data.  The scripts called appropriate \texttt{gLite} commands to retrieve the required data from the LFC, then ran the Locus Algorithm software (written in C) to generate the output for that task, which was again copied to the LFC for storage by gLite commands included in the grid job script.
      
In practice, this system was tested and evaluated using an evolutionary software development model, and a number of metrics were observed regarding performance of the system on the grid.  Note that a slow response cycle for grid jobs (minimum 5 minutes, maximum 3 days) meant that grid job testing emphasised factors impacting scalability below, while most testing was carried out in a serial computing environment. See \cite{grid_system_paper} for more information on test results and other performance metrics of the system.

This project used grid computing for three phases of operation: Parameterisation, discussed in Subsection \ref{parameterisation}, the API, discussed in Subsection \ref{API} and the Pipeline, discussed in Subsection \ref{Pipeline}.

\subsection{Grid Job Parameterisation}
\label{parameterisation}
The Parameterisation software is part of the job management system responsible for partitioning an overall task into jobs, each of which consists of a number of \textit{work units}.   These work units are the smallest unit of work which can be carried out by a given element of the software: they are different for each mode of operation.  For the API, which is responsible for extracting data from the Source Format into the Local Format, the work unit is a Source Catalogue file.  For the Pipeline, which determines the optimum pointing for a list of targets, the work unit is a single target when operating in Target List mode and a Local Catalogue file when operating in Catalogue traversal mode as detailed in in Subsection \ref{Pipeline}.  A job is a collection of work units bundled together such that they can be passed to the appropriate program and it will be able to carry out its task.  The number of work units per job is selected by the user when running the parameterisation software and is optimised to ensure the task will fit the processing requirements of the grid system as discussed in \citet{grid_system_paper}.

The Parameterisation software is designed to provide flexible input to the API and Pipeline programs.   Early prototyping on these programs revealed that they would become heavily dependent upon SDSS file and directory structures without a layer of data abstraction to allow this information to be passed in by means of a parameter file. The parameterisation software generates two types of binary parameter file, PRM files for the API, and PPR files for the Pipeline. These files contain explicit file paths, target lists and control variables for the API and Pipeline programs.  Because of this, the software can be applied to other catalogues with changes to the parameterisation software, but minimal changes to the other programs.

When developing the bash shell scripts used for grid management, it was determined that the PPR and PRM formats cannot easily be interpreted by the shell.  As a pragmatic solution to this problem, text files containing the file paths were implemented which are accessed by the shell scripts, as a supplement to the binary files used by the C programs.  These files, though stored with the file extension .txt, are referred to as Parameter Text (PRT) files in this paper to distinguish them from ordinary plain text files.

      \begin{figure}[!htb]
        \center{\includegraphics[width=0.5\textwidth]
        {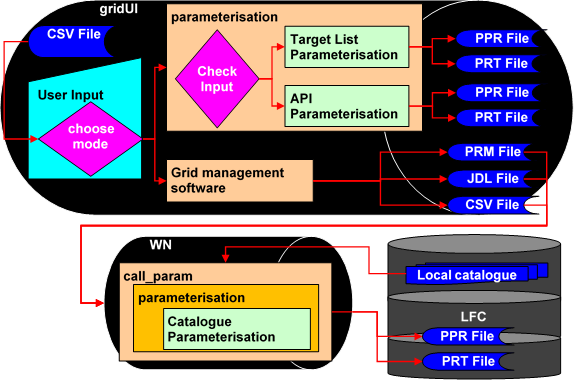}}
        \caption{\label{fig:parameterisation}Parameterisation Modes.  User input, together with the data contained in the CSV file from the CAS determines which mode parameterisation employs.  Target List and API parameterisation run on and store their output on gridUI, while Catalogue Parameterisation mode is a grid job, managed by its own suite of GMS, storing the parameter files generated on the LFC.  Copied from \citet{creaner2016thesis}}
      \end{figure}

The primary component of this system is a program called \texttt{parameterisation}.  This program takes as input a combination of command line arguments and data stored in CSV files of various structures .  The output of this program is a number of parameter files, the size and structure of which is determined by command line arguments which designate which mode of operation the program employs.  It operates in three modes: API, Target List Pipeline and Catalogue Traversal Pipeline as illustrated in Figure \ref{fig:parameterisation}.  Based on the input command line parameters and the contents of the CSV file, the program automatically determines which mode it will operate in.  

In all modes, \texttt{parameterisation} takes the following common arguments: the number of \textit{work units} to include in each grid job and three paths in which to: (1) store output parameter files (2) find the CSV file containing data to parameterise (3) find the FITS files the parameter file will refer to.    For both pipeline modes, the user must supply observational parameters regarding the telescope the output is intended for use with.  These parameters are FoV size (in degrees), Resolution (in degrees), Maximum magnitude difference (in magnitudes) and Maximum colour index difference (in magnitudes). Finally, to operate in Catalogue Traversal mode \texttt{parameterisation} requires a PRM file which contains the paths to the files for each of the target fields.  These parameters are summarised on Table \ref{table:param_args}.

\begin{table}[htb!]
\centering
\begin{tabular}{cccc}
\hline
   \textbf{Argument} & \textbf{API } & \makecell{\textbf{Pipeline}\\ \textbf{Target List}} &  \makecell{\textbf{Pipeline}\\ \textbf{Catalogue}} \\ \hline \hline
\textbf{\#{} Work Units} & \checkmark  &\checkmark    &\checkmark     \\ 
\textbf{Output Path} & \checkmark  &\checkmark    &\checkmark     \\ 
\textbf{CSV Path} & \checkmark  &\checkmark    &\checkmark     \\ 
\textbf{FITS Path} & \checkmark  &\checkmark    &\checkmark     \\ 
\textbf{FoV Size} & $\times$  &\checkmark    &\checkmark     \\ 
\textbf{Resolution} & $\times$  &\checkmark    &\checkmark     \\ 
\textbf{$\Delta$mag\textsubscript{max}}   & $\times$  &\checkmark    &\checkmark     \\ 
\textbf{$\Delta$col\textsubscript{max}}  & $\times$  &\checkmark    &\checkmark     \\ 
\textbf{PRM Path} & $\times$  &$\times$     &\checkmark     \\ \hline 
\end{tabular}

 \caption{List of the arguments to \texttt{parameterisation} and which modes each is used with.}
 \label{table:param_args} 
\end{table}

\subsubsection{Parameterisation in API mode}
\label{API_Parameterisation}

Input for the API mode has no additional arguments and the CSV file contains the four \textit{field identifiers} used in SDSS to identify fields which correspond to files in the catalogue: \texttt{run}, \texttt{rerun}, \texttt{camcol} and \texttt{field}.  When parameterise is called with the four arguments indicated on Table \ref{table:param_args}, it operates in API mode.

\begin{table}[htb!]
\begin{tabular}{c}
Naming Convention\\
\texttt{R/r/calibChunks/C/tsObj-RRRRRR-C-r-FFFF.fit} \\ \hline
Example\\
\texttt{1458/40/calibChunks/4/tsObj-001458-4-40-0352.fit}

\end{tabular}

 \caption{SDSS DAS File naming convention \citep{SDSSDAS}. Given Run (R), Rerun (r), Camcol (C) and Field (F). Rerun and Field in the filename are padded with leading zeroes to the length illustrated, but leading zeroes are not used in the directory names, nor are they used for Rerun or Camcol in either the filename or the directory. An arbitrarily selected example is shown. Copied from \citet{creaner2016thesis}}
 \label{table:SDSS_name} 
\end{table}

In API mode, \texttt{parameterisation} parses a list of field identifiers from SDSS into file names and paths according to the SDSS DAS directory stucture as shown on Table \ref{table:SDSS_name} and stores these paths into PRM files which contain the full paths to each of those files.  The internal structure of PRM files is duscussed in \citet{locus_software_paper}.

\subsubsection{Parameterisation in Target List Pipeline mode}
\label{param_Target}

 In Target List mode, information about the targets is passed in through a CSV file which contains the position and magnitutes of the targets, and the SDSS parameters as above which identify the fields in which stars may be found which can be included in a FoV with the target. To operate in Target List mode,  \texttt{parameterisation} must be passed 8 parameters as shown on Table \ref{table:param_args}.

In this mode, \texttt{parameterisation} generates PPR files.  In target list mode, each target has an individual set of fields around it, referred to as a \textit{mosaic} from the tiling pattern used to generate it.  For each mosaic, parameterise parses the list of field parameters given in the CSV file into filenames in the Local Catalogue which follow the same pattern as the Source Catalogue shown in Table \ref{table:SDSS_name}.  It also stores the Target position and magnitude so they can be used in the pipeline and associated with the correct mosaic. The internal structure of PPR files is duscussed in \citet{locus_software_paper}

\subsubsection{Parameterisation in Catalogue Traversal Pipeline mode}
\label{param_Catalogue}
In Catalogue Traversal mode, the CSV file contains a two lists of fields.  Firstly, it contains a list of fields in the catalogue such that the targets in those fields can be used as targets for the pipeline.  These are referred to as ``target fields.'' For each such target field, the CSV file also includes a list of fields in which any star in that field can be included in a FoV with any star in the target field.  These fields are referred to as reference fields.  Prior to running in Catalogue Traversal mode, the target fields must be passed to \texttt{parameterisation} in API mode to generate PPR files containing the paths to the files for the target fields.  This PRM file is then passed as a parameter to \texttt{parameterisation} as illustrated in Table \ref{table:param_args}.

In order to traverse an entire catalogue, and generate pointings and scores for every target in the catalogue, every file in the catalogue must be used as a target field. Therefore, the scale of the Catalogue traversal mode is much greater than the other two modes and thus, it has to be operated as a grid task in its own right.  As a result, the CSV file generated as output from the CAS is subdivided by the grid management software into a number of smaller CSV files.  These smaller CSV files are used as input to each of the parameterise grid jobs in turn as described in Subsection \ref{jobscripts}.

Each grid parameterisation job takes many files from the LFC, and generates PPR files containing target lists and the paths corresponding to the reference field for each set of targets.  These PPR files are then copied to the LFC for use in the pipeline when it is operating in catalogue traversal mode as discussed in brief in Subsection \ref{Pipeline} and in more detail in \citet{locus_software_paper}.

\subsection{Grid Job Scripts}
\label{jobscripts}
The grid was used for four operations within the scope of the project.  Extracting data from the source Catalogue to the Local Catalogue through the API, calculating pointings for a list of quasar targets to generate the quasar catalogue as shown in \citet{quasarpaper}, generating PPR files for catalogue traversal as discussed in Subsection \ref{param_Catalogue} and calculating pointings for all targets in the Local Catalogue as required for \citet{ZenodoXOPCatalogue}.    

      \begin{figure}[!htb]
        \center{\includegraphics[width=0.5\textwidth]
        {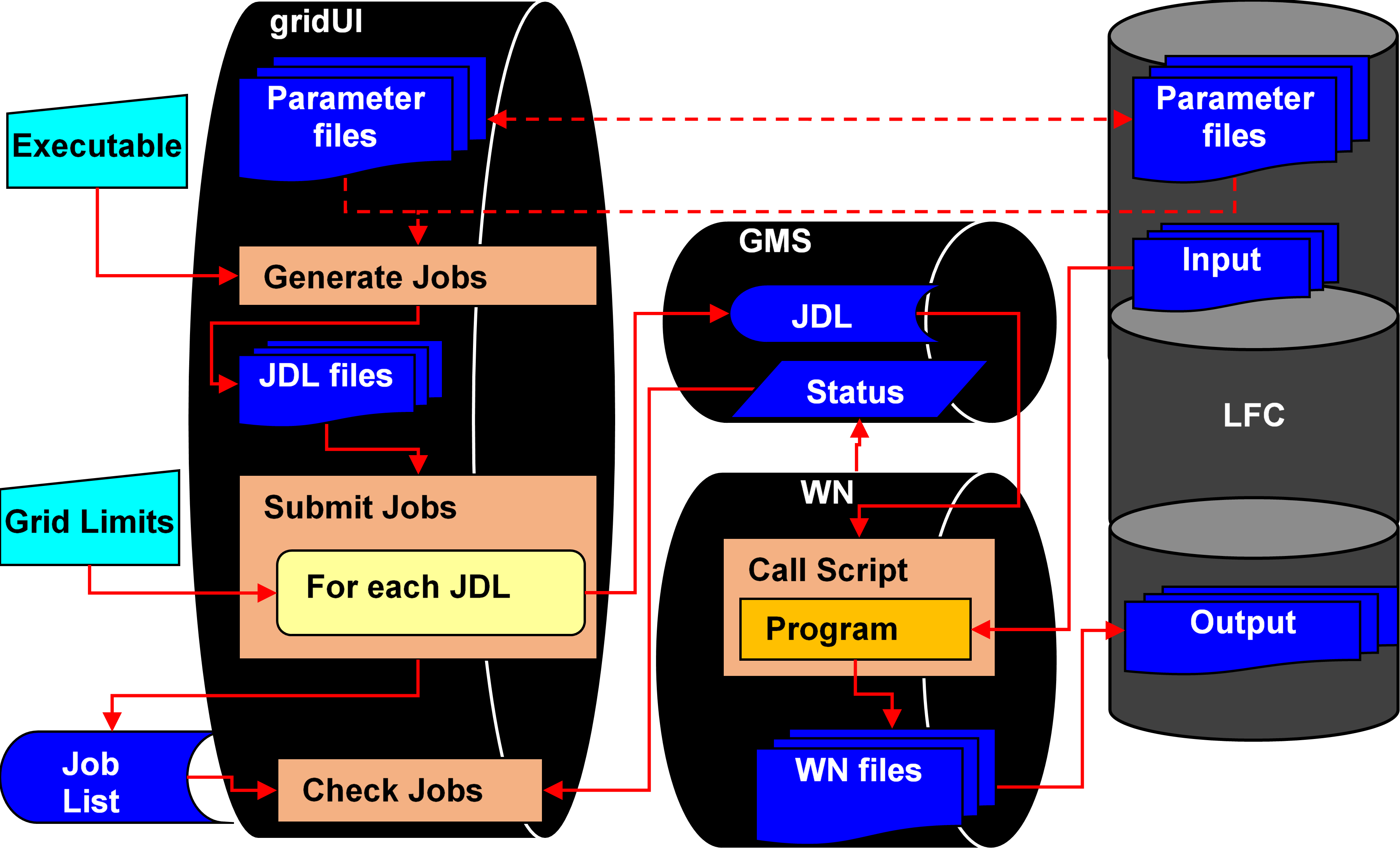}}
        \caption{\label{fig:job_scripts} Schematic structure of grid managment scripts.  Copied from \citet{creaner2016thesis}}
      \end{figure}
  
All grid operations require grid management software, written in Bash to operate.  For each of these operations, the same four components of the Grid Management Software (GMS) exist: job generation (see Subsection \ref{job_generation}), job submission (see Subsection \ref{job_submission}), job calling (see Subsection \ref{job_calling}) and job monitoring (see Subsection \ref{job_monitoring}). The design of these components is largely identical across each of the four operations.  The differences are treated here as minor variations to the design and highlighted where significant. 

\subsubsection{job generation}
\label{job_generation}
The \texttt{generate\textunderscore{}jobs} scripts operate on gridUI.  They use existing Parameter files generated by the \texttt{parameterise} program in one of its modes to create a series of JDL files, each of which is submitted as a grid job.  Parameter files are stored on gridUI except in the case of the Pipeline in Catalogue Traversal Mode, in which case the Parameter files are on the LFC.  These files are listed (using \texttt{ls} or \texttt{lfc-ls} commands depending on the location of the parameter files) and the output of that list is piped by script into a template using scripting variables.  The output of these scripts is a directory containing a set of JDL files which were later submitted by \texttt{submit\textunderscore{}jobs}.

\subsubsection{job submission}
\label{job_submission}
The \texttt{submit\textunderscore{}jobs} scripts run on gridUI and submit a set of JDL files from a location specified by the user to the JSS in a controlled manner.  \texttt{submit\textunderscore{}jobs} takes, as an argument, the user-defined limit to how many grid jobs NGI permitted that user to have running simultaneously.  It then submits jobs in batches of 10 at a time from gridUI, waiting to ensure that all jobs previously submitted have started running before submitting any more.  It also monitors how many jobs are running at a time, stored as the variable \texttt{running\textunderscore{}job\textunderscore{}count}.  If (\texttt{running\textunderscore{}job\textunderscore{}count + 10}) is greater than the limit of simultaneous jobs, submission is halted until \texttt{running\textunderscore{}job\textunderscore{}count} is low enough that submitting a batch of new jobs would not cause it to exceed the limit.  The output from this set of scripts is a text file containing a set of job identifiers which were used by \texttt{check\textunderscore{}jobs} to monitor the progress of those grid jobs.

\subsubsection{job calling}
\label{job_calling}
The job calling scripts are specified in the JDL files and when the JSS assigns a WN to carry out a grid job, these are the scripts that are called.  The arguments to the scripts are specified in the JDL file which is what determines the difference between each individual grid job.  The job may also be supplied with some files from \texttt{gridUI} which may or may not be compressed, depending on the particular operation.  The \texttt{call\textunderscore{}job} scripts are more different between the different operations than the other grid jobs scripts, but each of these scripts has the following common features.  

Firstly, the script extracts any compressed files provided by the JSS from \texttt{gridUI} into the working directory of the WN.  It then creates a directory structure that will hold the executable files which will carry out the main task.  Next, the script copies the main program from the LFC to the WN and makes it executable.  If it was not supplied through the JSS, the parameter file is then copied from the LFC to the WN.  Finally, by iterating through the lines of the text parameter file, the script creates the directory structure required and copies the listed fits files from the LFC to the WN into that location. 

With the programs and data now located on the WN, the main task of the grid job is executed by calling the main program or programs with appropriate arguments (supplied in the JDL file to the JSS).  These programs are typically called in low-verbosity modes or with their output redirected to \texttt{/dev/null} as grid jobs are run non-interactively and in NGI there was limited space for text diagnostic output from grid jobs. These programs generate output in pre-determined directories in the working directory of the WN. 

For later use (in the case of the API and parameterisation tasks) and for output and publication (in the case of pipeline tasks) the data must be moved from the WN to the LFC.  This task is accomplished by creating the appropriate directory structure on the LFC, then copying the output file(s) from the WN to the LFC.

\subsubsection{job monitoring}
\label{job_monitoring}
Finally, \texttt{check\textunderscore{}jobs} runs on \texttt{gridUI} and interfaces with the GMS to track the status of on-going grid jobs.  The job lists output from \texttt{submit\textunderscore{}jobs} is used as an input to \texttt{check\textunderscore{}jobs}.  It generates a list of job statuses which can be piped to a file or displayed on \texttt{std.out} as determined by the user.  These statuses can also be piped through standard Unix tools such as \texttt{grep} and \texttt{wc} to isolate individual jobs or particular groups of job statuses.

\section{Core Programs}
\label{Core}

The programs that form the core of this system are the API (which extracts data from the SDSS format to a reduced format known as the Local Catalogue format) and the Pipeline (which calculates optimal pointings for a set of targets).  These programs are designed to work independent of their environment and are thus portable between different implementations of the system.  Thus, the programs are not grid-native and they require the wrapper of the grid scripts as outlined in Subsection \ref{jobscripts}.  The details of the design of these systems is discussed in \citet{locus_software_paper}, together with the design requirements, constraints and implementation of the system in isolation from the grid.  

In the scope of this paper, it is essential to know how the requirements for this software impact the grid system design, and how the required inputs and outputs of the core system influence the design of the grid system.  These programs, which are written in C, and make extensive use of the CFITSIO Library by \citet{pence1999cfitsio} are described in brief below.

\subsection{SDSS data ingestion API}
\label{API}
The Locus Algorithm is designed to be flexible with regards to source data.  By extracting data from a Source Catalogue into a Local Catalogue, the data required for the algorithm is retained, while other data is discarded.  The Local Catalogue format was designed such that data from any catalogue containing position (RA, Dec) and magnitude information could be extracted into that format and used with the Locus Algorithm.

The source catalogue for this project was the SDSS Catalogue of Calibrated Objects.  This catalogue was held on the SDSS Data Archive Server (DAS) \citep{SDSSDAS} in a collection of many Calibrated Objects files (\texttt{tsObj} files).  This catalogue was downloaded to the LFC and stored there for processing.  These files each contain 146 columns, and each row refers to an observation of an object by an SDSS camera. Only a few of the attributes recorded in those 146 columns are required for the software system presented here. Many of these records are of non-stellar objects not suitable for use as reference stars.  In addition, due to the observation pattern used by SDSS, many of these records are not ``primary'' observations as defined in the SDSS Image processing flags \citep{SDSScleansample}.

      \begin{figure}[!htb]
        \center{\includegraphics[width=0.47\textwidth]
        {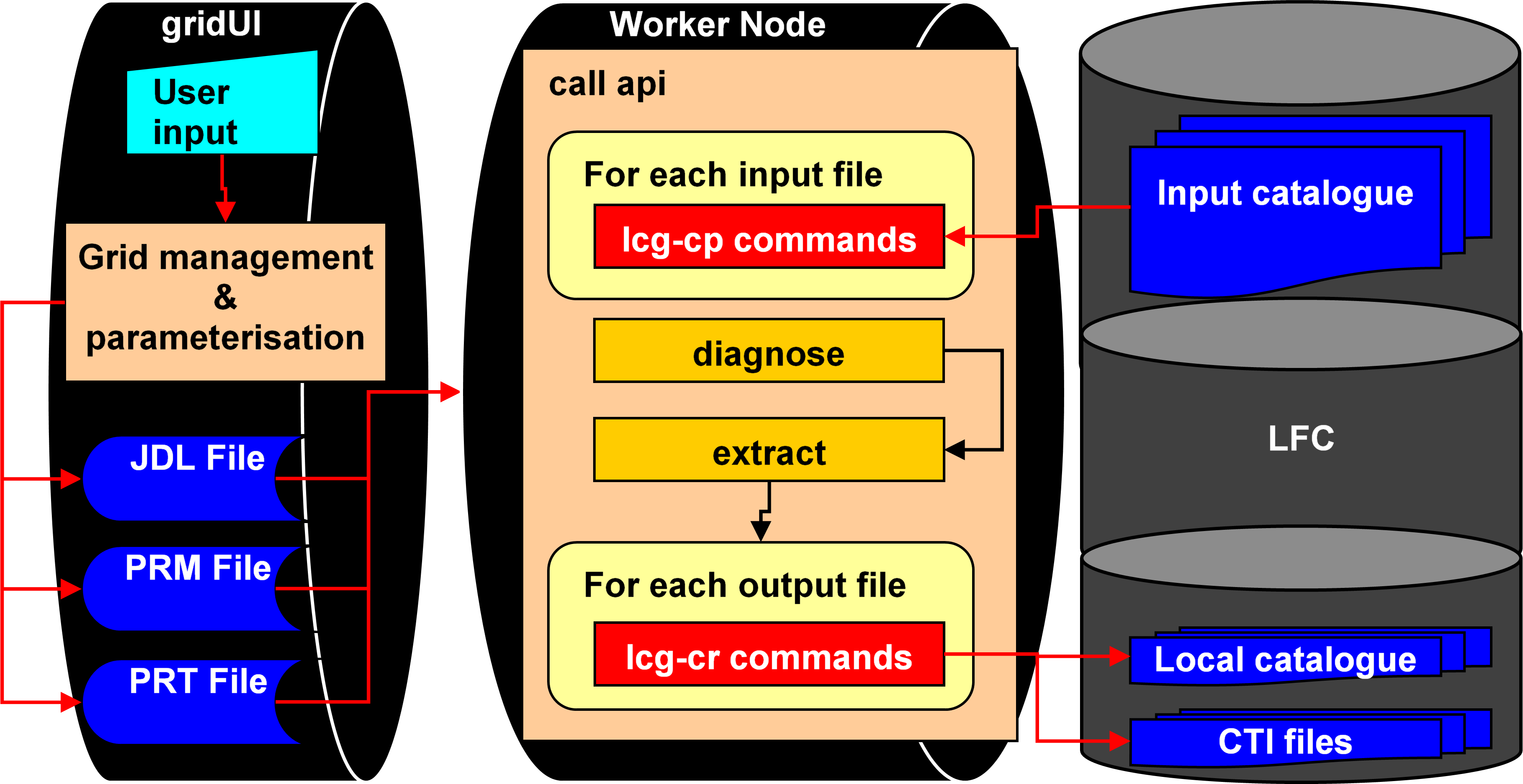}}
        \caption{\label{fig:api} Structure of the SDSS Data ingestion API.  Copied from \citet{creaner2016thesis}}
      \end{figure}

The SDSS Data Access API is designed to extract data from the SDSS catalogue in such a way as to minimise the data volume that is accessed in the pipeline by both projection and selection operations on the source catalogue files. The selection operation is carried out by filtering to only those records which meet the SDSS clean sample of stars criteria, which discards non-stellar and non-primary sources \citep{SDSScleansample}.  The projection operation consists of selecting only the position (RA, Dec) and magnitude (\textit{ugriz}) columns from the data as the other columns are not required for the operation of the Locus Algorithm \citep{locuspaper}.

The API is split into three conceptual blocks as illustrated in Figure \ref{fig:api}: work carried out on \texttt{gridUI}, work carried out on the WN, and interactions with the LFC.  

The first block consists of parameterisation as described in Subsection \ref{API_Parameterisation}, where user input is given to the parameterisation system to determine how many work units (SDSS fields) to assign to each grid job, which are then listed and stored in paramter files on \texttt{gridUI}.  For each of these parameter files, a JDL file is generated by the \texttt{generate\textunderscore{}jobs\textunderscore{}api} script which creates a series of \texttt{.jdl} files.  These files are submitted to the JSS, which assigns each job to a WN.

\texttt{call\textunderscore{}api} operates on a WN, and takes two parameter files as inputs from gridUI. One is in PRT text format and one is in PRM binary format.  The PRT file specifies the LFN paths in the LFC to a set of input SDSS fits files which are to be processed by the API to create the corresponding local catalogue files and is used by the shell script.  A for loop iterates through this file, and uses the lcg-cp command to copy the listed files from the LFC to the WN.  

Two programs, Diagnose and Extract, are then copied from the LFC to the WN.  These programs use the PRM parameter file as an input to access the \texttt{fits} files.  The Diagnose program accesses and analyses the data contained within the input files, and identifies the columns in the data table.  The column names together with information regarding their structure are then stored in a file. Next, the Extract program carries out the substantive work of the API.  Taking the columns identified by Diagnose, it identifies the columns containing only the relevant data: Right Ascension, Declination and Magnitude (itself an array of 5 double values).  It then applies  the SDSS clean sample of stars algorithm to exclude entries which are not primary entries for stars\citep{SDSScleansample}.  The data for the remaining entries in those three relevant columns for this project are then written to a file which forms part of the local catalogue.  

Finally, \texttt{call\textunderscore{}api} then uses \texttt{lcg-cr} to copy all of the Local Catalogue files to the LFC for storage.  These files are then used as the inputs to the Pipeline

\subsection{Pipeline}
\label{Pipeline}

\begin{figure}[!htb]
        \center{\includegraphics[width=0.47\textwidth]
        {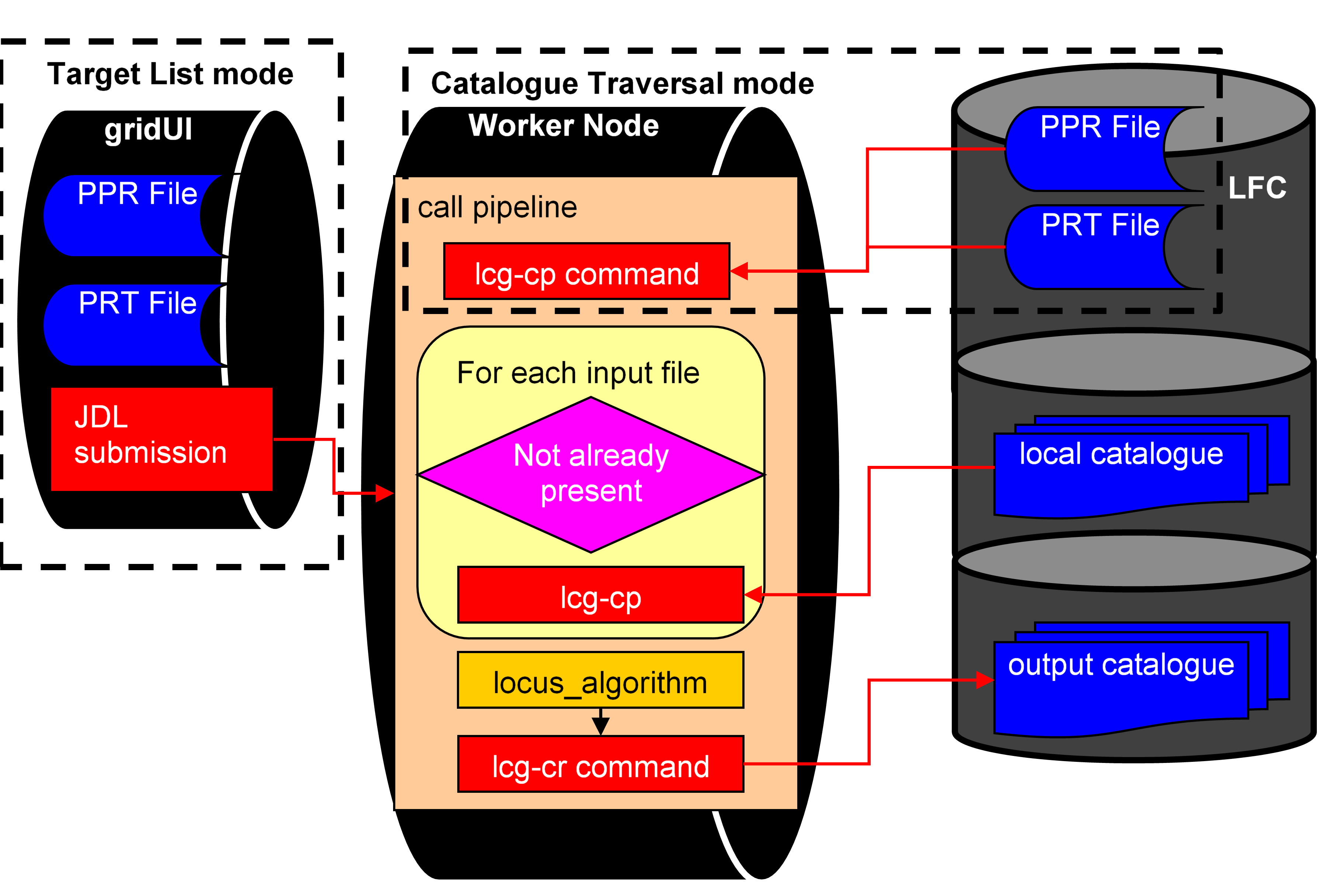}}
        \caption{\label{fig:pipeline} Structure of the SDSS Data ingestion API.  Copied from \citet{creaner2016thesis}}
      \end{figure}

The Data Pipeline applies the Locus Algorithm as defined in \citet{locuspaper} to a set of targets to produce output files which form the Output Catalogue(s).  It operates in two modes: Target List and Catalogue Traversal. Target list mode is used when a set of one or more targets are selected in advance by the user and are submitted to the pipeline for processing.  This mode was used to produce the Quasar Catalogue as discussed in \citet{quasarpaper}.  Catalogue traversal mode uses an existing catalogue or subset of a catalogue to produce the target list. This target list is then submitted to the pipeline in a similar way to the Target list mode discussed above.  This mode was used to produce the Exoplanet Catalogue as presented in \citet{ZenodoXOPCatalogue}.  The structure of these modes is outlined in Figure \ref{fig:pipeline}.

Much of the software is designed to ignore the distinction between these two modes, for example, by treating a single target as a list of length one.  Distinctions between operational modes are therefore discussed only as needed, as shown in Figure \ref{fig:pipeline}, where dashed boxes are used to indicate modules that differ between the two modes. 

In Target List mode, a target list consisting of the positions (RA, Dec) of a set of specific targets (e.g. quasars) is submitted through an SQL script to the CAS \citep{SDSSCAS} together with FoV size.  For each target in the list, the script identifies which fields include any objects which are inside the Candidate Zone (CZ) as defined in \citet{locuspaper}.  The CAS returns a list of field identifiers as required for parameterisation as described in Subsection \ref{param_Target}.  This list of identifiers is combined with observational parameters and passed through the \texttt{parameterisation} and \texttt{generate\textunderscore{}jobs} scripts as defined in Subsections \ref{param_Target} and \ref{job_generation} to generate a set of grid jobs. 

In Catalogue Traversal mode, a field list, consisting of the field identifiers for all fields in the catalogue is submitted to the CAS instead of a target list.  The SQL script identifies, for each target field, the neighbouring fields which have at least one object within the CZ for any target in the target field.  The target fields and reference fields are then passed through the \texttt{parameterisation} and \texttt{generate\textunderscore{}jobs} scripts as above.

Each grid job calls a script named \texttt{call\textunderscore{}pipeline} which uses the parameter files to identify the necessary local catalogue files and generates the output catalogue based on the listed targets. Two versions of the Call Pipeline script execute the pipeline, one in each of the two modes.  The primary distinction between these modes is that when run in Target List mode, the Parameter files are stored on gridUI, while in Catalogue Traversal mode those files are stored in the LFC.  This means that in the first case, the job submission process submits the PPR files as part of the grid job, while in the second, they have to be copied out of the LFC using gLite commands.  The two versions are otherwise identical as shown in Figure \ref{fig:pipeline}.

For each Local Catalogue file listed in the PRT file, \texttt{call\textunderscore{}pipeline} checks to see if the file is already present on the WN, and if not, copies it from the LFC to the WN.  Note that each file is likely to be used by more than one target in the target list, especially in Catalogue Traversal mode.  

Call pipeline then executes the C program \texttt{locus\textunderscore{}algorithm}, defined in greater detail in \citet{locus_software_paper}.  This program  with the colour argument provided in the JDL file, redirecting its output to /dev/null unless operating in verbose mode for debugging purposes.  This program implements the Locus Algorithm and generates an output file. 

Finally, the output file is copied and registered to the LFC and any errors are reported to the Grid Management Software.

\section{Data Structures and Management}
\label{Data}

Data concerns are central to the design of this system.  Any implementation of the system would require the manipulation of large volumes of data from a catalogue.  Since data transfer operations are generally much slower than data processing operations, a design constraint was to minimise these operations wherever possible.  In addition, a design goal for this project was to develop software that could be flexible as to the source catalogue that it operates with, such that the system can be used with future catalogues such as Gaia or LSST as and when they become available.  As a result of both of these constraints, the system was designed with data abstraction as discussed in Subsection \ref{abstraction}.  

The system was required to be flexible enough to handle input data from multiple sources, and produce output date in such a way as to be accessible by users with different requirements.  The mandated the use of several internally and externally designed data types.  The design requirements and constraints for these data types are discussed in detail in \citet{locus_software_paper} and summarised in Subsection \ref{datatypes}.

The physical and logical structure of the grid influenced the data storage structure and the means used to transfer data between them.  This is detailed in Subsection \ref{datasystems}.

\subsection{Data Abstraction}
\label{abstraction}
While the system was implemented with the SDSS Catalogue, it was developed with flexibility as to the source of the data.  This flexibility was enabled by the use of the Data extraction API described in Subsection \ref{API}.  This system extracted the data that was required for the algorithm from the source catalogue and processed it into a new structure called the Local Catalogue.  The Local Catalogue consisted of a series of FITS files which could be read by the main data pipeline.  

In future iterations of this system, it is envisaged that a new API could be developed to read data from the structure of other catalogues (e.g. Gaia) and write it to the Local Catalogue format.  This abstraction provides a layer of source-independence for the processing pipeline

As a practical matter, the directory structure of the local catalogue mimicked the structure of the source catalogue.  However, the local catalogue directory structure was abstracted from the pipeline software through the parametrisation software as shown in Subsection \ref{parameterisation}.  This software queried the CAS to identify targets (or target fields) and the reference fields that would be needed, and provided the pipeline with the paths to those fields.  This parameterisation software can be updated in a modular fashion to allow it to work with new catalogues.

Because of these abstractions, the core software of the pipeline is independent of the structure of the source catalogue, depending instead only on the local catalogue.

\subsection{Data Types}
\label{datatypes}

Data in this project was stored in a variety of formats to fit with the requirements of the sources and the systems that were employed.  These are summarised in Table \ref{table:data_types}, and discussed in greater detail in \citet{locus_software_paper}.

Data in the Source Catalogue was in the Flexible Image Transport System (FITS) format, a widely used format in the astronomical community.  To minimise the number of dependencies of the software system, it was decided that the Local Catalogue and the Output Catalogue would also be stored in FITS format.

\begin{table}[htb!]
\centering
\begin{tabular}{c p{5cm} c}
\hline
   \textbf{Data Type} & \textbf{Used in } & \textbf{Novel}\\ \hline \hline
\textbf{FITS} & Source Catalogue, Local Catalogue \&{} Output Catalogue  &$\times$     \\ 
\textbf{CSV} & Parameterisation  \&{} Output Catalogues &$\times$     \\ 
\textbf{JDL} & Grid Jobs  &$\times$     \\ 
\textbf{CIT} & API  &\checkmark     \\
\textbf{PRM} & API \&{} Pipeline  &\checkmark     \\ 
\textbf{PPR} & Pipeline   &\checkmark     \\  
\textbf{PRT} & Grid Jobs  &\checkmark     \\ \hline 
\end{tabular}

 \caption{Table of the data types used in this project, in which project components they are used and whether they were developed for this project or not.}
 \label{table:data_types} 
\end{table}

The parameterisation software which formed part of the job management and data abstraction elements produced data in two novel formats developed for this project.  PRM files were used to identify files in the source or local catalogue without structure for grouping.  PPR files were used to identify targets for processing in the pipeline, grouped together with files to be used in groups to form mosaics around those target(s).  PRM and PPR files are binary format files which are interpreted by the core software.  PRT files are text files including the lists of target files from PPR or PRM files which can be interpreted by the shell scripts used for grid job management as discussed in Subsection \ref{jobmanagment}.

CTI (Catalogue information) is a file format which was built for this project to enable information about the structure of catalogue files to be stored in a lightweight, accessible format as part of the API as discussed in \cite{locus_software_paper}.

The Job Description Language (JDL) is used to define grid jobs as part of the \texttt{glite} system \citep{glite}.  JDL files are automatically created by the \texttt{generate\textunderscore{}jobs} scripts which access the list of parameter files for a given grid operation and create new JDL files which can be submitted to the JSS as discussed in Subsection \ref{jobmanagment}.

Comma Separated Value (CSV) files are used in two ways in the project.  First the outputs from the CAS are delivered in this format.  The CAS supplies the lists of fields in the source catalogue which are used in the parameterisation software to define the paths to the files used in the API.  The CAS also supplies lists of targets and their associated fields which are used by the parameterisation software to generate the jobs for the pipeline.  In addition, the output catalogues, which were created in FITS file format, have also been converted into CSV format for publication, for the benefit of users who do not use FITS.

\subsection{Data Storage Systems}
\label{datasystems}

\begin{figure}[!htb]
        \center{\includegraphics[width=0.47\textwidth]
        {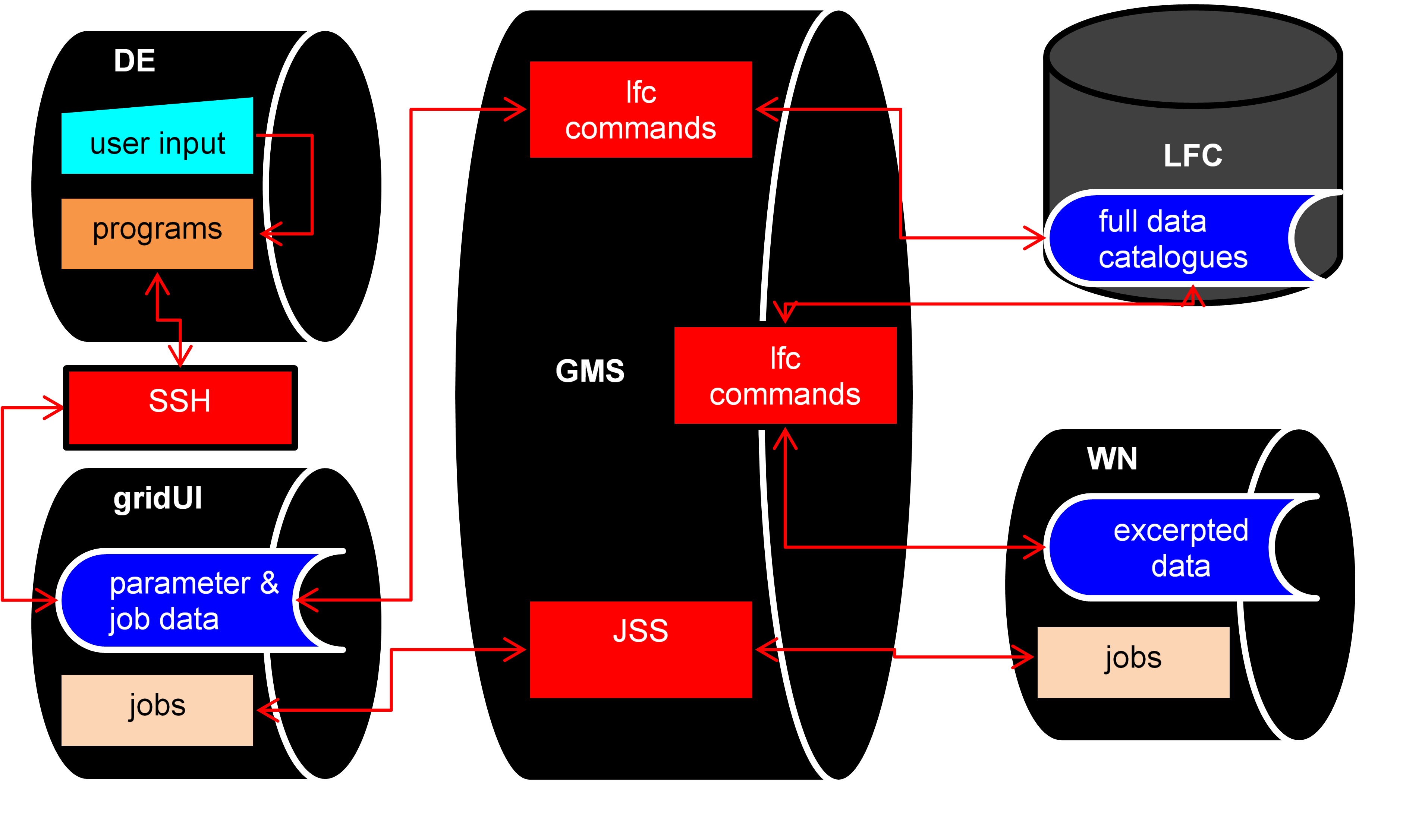}}
        \caption{\label{fig:data_storage} Overall layout of the grid data storage structure.  The software was developed in the Development Environment (DE).  Programs and user input were copied from the DE to gridUI through Secure SHell (SSH).  The GMS handled data and software transfer between \texttt{gridUI}, the LFC and the WNs using the \texttt{lfc-*} commands of the \texttt{glite} system.  Copied from \citet{creaner2016thesis}}
      \end{figure}

Over the course of this project, data was created, stored and used on a variety of different storage elements, some of which were volatile and others were persistent relative to the timeline of the project.  Data transfer between these elements was a key design consideration, as network data transfer operations are slow by comparison to local file I/O operations.  A schematic of these data storage elements is presented in Figure \ref{fig:data_storage}

All user interaction in the project was carried out on the Development Environment (DE).  This was a local Windows-based system with virtual environments configured to reflect WNs.  This allowed for rapid development and testing of software independent of grid operations.  This system was connected to \texttt{gridUI} through Secure SHell (SSH) which allowed both file transfer (for software) and user input to be communicated to the grid.

As described in Subsection \ref{gridui}, \texttt{gridUI} was the gateway to the grid and had access similar to that of a WN.  Programs, parameter files and grid jobs were staged here for submission to the grid.  The \texttt{glite} system was used to submit these elements.  Programs were transferred to the LFC through the GMS by submitting \texttt{lfc-cr} commands.  JDL files were created on \texttt{gridUI} and submitted to the JSS.  The JSS would then assign the job described in the JDL file to an available WN and monitor its progress.  The user could submit requests to the JSS using the job monitoring software described in Subsection \ref{job_monitoring} to track the progress of a grid operation.

As described in Subsection \ref{job_calling}, grid jobs typically included \texttt{lfc-cp} commands to transfer programs and the required data from the LFC to the WN to allow a job to begin, and \texttt{lfc-cr} commands to transfer output data to the LFC when the job was complete.  The data abstraction described in \ref{abstraction} above greatly reduced the volume of data to be copied from the LFC to the WN.  

\section{Practical Implementation}
\label{Implementation}

In practice, a number of issues arose with this project which impacted its performance on the grid.  The I/O dominated nature of this project meant that it had different performance characteristics to many previous tasks which were processed using grid computing.   

As a matter of policy by the grid managers, jobs on NGI were restricted to a maximum of three days runtime, after which point they would be automatically terminated.  As a result, when allocating work to a grid job, it was necessary to ensure that the amount of data selected was such that the job would be completed within this limit.  

Different uses of the software required different work units, each of which had different requirements and run times.  A work unit is defined as the smallest element of the data that can be processed in a grid job. Local unit testing, combined with small-scale grid testing was used to establish the runtime for each work unit.  When grid jobs were generated using the parameterisation software, the number of work units was input as a command line option as shown in Figure \ref{fig:grid_jobs}.  The number of work units was chosen by the user and submitted to the Grid Managment software (Figure \ref{fig:grid_management}) such that the run time was well within the limits.  Extended discussion of the performance metrics of this system are available at \cite{grid_system_paper}

\section{Discussion}
\label{Discussion}

Completion of this project highlighted a number of issues with the use of distributed computing systems, specifically gLite, for astronomical operations.  In addition, the metrics created during the project highlight distinct use cases for systems of this nature which should be considered in future projects.  These issues and use cases are discussed below.

One major issue which arose during this project was the suitability of the gLite LFC system to large volumes of simultaneous access requests.  Because the gLite system is designed with data security in mind, it activates separate authentication processes for each file accessed from the LFC \citep{glite}. This means that each file takes of the order of one second to access and download.  For typical grid problems where the data is contained in few large files, this issue is not significant.  In this case, where each job may access thousands of small files, this access time can become dominant.  The solution to this problem was to provide storage space at a shared network drive (NFS) which lacked the thorough security and authentication process of the LFC.  This allowed for faster data access for grid jobs using that system.

It is typical for grid job to produce log files which include the standard outputs.  These log files are typically small, of the order of kilobytes.  In this project, development versions of the system had produced verbose outputs to the screen containing much of the data from input files and the final resulting file.  If stored, this output would be comparable to or larger than the input data, amounting to many megabytes per grid job.  Log file storage systems were unable to manage this data volume and velocity, and as a result a silent mode was developed for the software which discarded much of the data which would otherwise have been sent to std.out and thus to log files.

While the software for this project was structurally divided into two components: the API and the Pipeline, these components are not fundamentally different in terms of how they used the grid: each read in the data from the LFC (Input), processed the data through some software (processing) and produced the output to the LFC (Output).  These steps are present in each of these and many other applications.  A distinct use case emerges however between the generation of the Local Catalogue and Quasar Catalogue on one side, and the Exoplanet Catalogue on the other.  

The former case is dominated by Data I/O operations especially when the gLite LFC is accessed and requires authentication for each file accessed. The latter case is balanced between Data I/O and Processing.  Finally, one can imagine a case where Processing is dominant, as is commonly the case with HPC applications.  Since the same system (the pipeline) demonstrates different characteristics dependent upon the inputs it is provided with, it follows that a simple examination of the code or algorithm design may not always be indicative of the characteristics of the system overall.

In future applications of this or related systems, therefore, it becomes important to distinguish not only the characteristics of the system, for example by analysis of algorithmic complexity, but also to test those system characteristics when applied to data of the type that is to be processed.

\section{Conclusions}
\label{Conclusions}

The objective of this project was to generate a system to analyse and identify optimal pointings for differential photometry for two sets of targets: Quasars (from SDSS Quasar Data Release 4) and stars which might be potential candidates to host exoplanets.  In both cases, optimal conditions for differential photometry were defined such that the target had a maximum number and quality of reference stars as defined in \citet{locuspaper}

To achieve this objective, two major software components were developed, and used to generate three catalogues, the Local Catalogue, the Quasar Catalogue and the Exoplanet Catalogue.  The Local Catalogue was used as an intermediate step in the generation of the two Output Catalogues.  Generation of these catalogues was completed using the gLite system of NGI.  Operational metrics on these systems showed a disparity in performance not between the software systems but rather between the application of the systems to different datasets.  

This project further demonstrates a the practical application of High Performance Computing to an astronomical data analytics problem.  From the metrics and practical observations, it is apparent that simple calculations of algorithmic complexity may not be sufficient to identify the dominant component in a project.  Rather, this system demonstrates the value of experimental assessment of system performance under operational conditions.

The software used in this project is available at on Github at \citet{githubrepo} and a paper describing the performance of this system is available at \cite{grid_system_paper}.  

\section*{Acknowledgements}
\textbf{Funding for this work}: This publication has received funding from Higher Education Authority Technological Sector Research Fund and the Institute of Technology, Tallaght, Dublin Continuation Fund (now Tallaght Campus, Technological University Dublin).

\textbf{SDSS Acknowledgement}: This paper makes use of data from the Sloan Digital Sky Survey (SDSS).  Funding for the SDSS and SDSS-II has been provided by the Alfred P. Sloan Foundation, the Participating Institutions, the National Science Foundation, the U.S. Department of Energy, the National Aeronautics and Space Administration, the Japanese Monbukagakusho, the Max Planck Society, and the Higher Education Funding Council for England. The SDSS Web Site is http://www.sdss.org/.

The SDSS is managed by the Astrophysical Research Consortium for the Participating Institutions. The Participating Institutions are the American Museum of Natural History, Astrophysical Institute Potsdam, University of Basel, University of Cambridge, Case Western Reserve University, University of Chicago, Drexel University, Fermilab, the Institute for Advanced Study, the Japan Participation Group, Johns Hopkins University, the Joint Institute for Nuclear Astrophysics, the Kavli Institute for Particle Astrophysics and Cosmology, the Korean Scientist Group, the Chinese Academy of Sciences (LAMOST), Los Alamos National Laboratory, the Max-Planck-Institute for Astronomy (MPIA), the Max-Planck-Institute for Astrophysics (MPA), New Mexico State University, Ohio State University, University of Pittsburgh, University of Portsmouth, Princeton University, the United States Naval Observatory, and the University of Washington.

This paper makes use of \textbf{CFITSIO}: A FITS File Subroutine Library by \citet{pence1999cfitsio}




\bibliographystyle{elsarticle-harv}
\bibliography{grid_system_paper}







\end{document}